\documentclass[aps,prl,reprint,groupedaddress]{revtex4-1}
\usepackage{graphicx} 
\usepackage{dcolumn} 
\usepackage{bm}
\usepackage{hyperref}

\begin{document}

\title{In-medium and isospin effects on particle production near threshold energies in heavy-ion collisions}
\author{Zhao-Qing Feng$^{1,2}$}
\email{Corresponding author: fengzhq@impcas.ac.cn}

\author{Wen-Jie Xie$^{1}$}

\author{Peng-Hui Chen$^{1,3}$}

\author{Jie Chen$^{1}$}

\author{Gen-Ming Jin$^{1}$}

\affiliation{$^{1}$Institute of Modern Physics, Chinese Academy of Sciences, Lanzhou 730000, People's Republic of China            \\
$^{2}$State Key Laboratory of Theoretical Physics and Kavli Institute for Theoretical Physics China, Chinese Academy of Sciences, Beijing 100190, People's Republic of China       \\
$^{3}$School of Nuclear Science and Technology, Lanzhou University, Lanzhou 730000, People's Republic of China}

\date{\today}
\begin{abstract}
Dynamics of pseudoscalar mesons ($\pi$, $\eta$, $K$ and $\overline{K}$ ) and hyperons ($\Lambda$ and $\Sigma$) produced in heavy-ion collisions near threshold energies has been investigated within the Lanzhou quantum molecular dynamics (LQMD) transport model. The in-medium modifications on particle production in dense nuclear matter are implemented in the model through corrections on the elementary cross sections and by inclusion of the meson (hyperon)-nucleon potentials, in which the isospin effects are considered. It is found that the transportation of particles are influenced with the in-medium corrections. The total number of pions is reduced with an isospin, density and momentum dependent pion-nucleon potential. However, the ratios of charged pions is enhanced with inclusion of the potential. The production of eta in the domain of mid-rapidities and high momenta is sensitive to the $\eta$-nucleon potential, but weakly depends on symmetry energy. The attractive antikaon-nucleon potential enhances the subthreshold $\overline{K}$ production and also influences the structure of phase-space distributions. Dynamics of etas, kaons, antikaons and hyperons is also influenced by the pion potential because of collisions between pions and nucleons (resonances). The impacts of mean-field potentials on particle dynamics are investigated, such as the phase-space distributions from rapidity and transverse momentum spectra, inclusive invariant spectra, collective flows etc.

\begin{description}
\item[PACS number(s)]
21.65.Ef, 21.65.Jk, 24.10.Jv
\end{description}
\end{abstract}

\maketitle

\section{I. Introduction}

Properties of hadrons in nuclear medium is interest in studying the Quantum Chromodynamics (QCD) structure in dense matter, in particular related to the chiral symmetry restoration, phase-transition from quark-glue plasma to hadrons, dynamics of hypernucleus formation, nuclear equation of state (EoS) etc \cite{Gi95,Fr07,To11}. High energy heavy-ion collisions in terrestrial laboratory provide a unique possibility to study the in-medium properties of hadrons in dense nuclear matter and to extract the high-density behavior of the nuclear symmetry energy (isospin asymmetric part of EoS). It has obtained progress in extracting the in-medium properties of hadrons in dense nuclear matter, in particular for strange particles $K$, $\overline{K}$, $\Lambda$ and $\Sigma$ \cite{Li95,Li97,Ca99,Fu06,Ha12}.

Dynamics of particles produced in heavy-ion collisions near threshold energies is a complicated process, in which the production and transportation in dense nuclear medium would be different in comparison to the in-vacuum cases. Density and momentum dependent potentials have to be implemented in correctly understanding the phase-space distributions of particles. Furthermore, dynamics of isospin particles could be modified by the mean-field potentials in constraining the high-density information of symmetry energy. Besides nucleonic observables \cite{Li00}, particles produced in heavy-ion collisions would be preferable probes for extracting the information of high-density phase diagram. Kaons as probing the high-density EoS were proposed for the first time \cite{Ai85}. The available experimental data from KaoS collaboration for $K^{+}$ production favored a soft EoS at high baryon densities associated with transport model calculations \cite{St01,Li95b,Fu01,Ha06,Fe11}. Similar structure for $\Lambda$ production on the EoS was found in Ref.\cite{Fe11}. The ratios of isospin particles produced in heavy-ion collisions such as $\pi^{-}/\pi^{+}$, $K^{0}/K^{+}$, $\Sigma^{-}/\Sigma^{+}$ etc \cite{Li02,Li05,Fe06,To10,Pr10,Fe10}, neutral particles such hard photons, $\eta$ etc \cite{Yo08}, and the flow difference of isospin particles \cite{Gi10,Fe12} have been proposed as sensitive probes for extracting the high-density behavior of the nuclear symmetry energy (isospin asymmetric part of EoS). Although different interpretations on the $\pi^{-}/\pi^{+}$ ratios are concluded in constraining the high-density symmetry energy with transport models \cite{Xi09,Fe10b,Xi13} in combination with the experimental data from the FOPI collaboration \cite{Re07}. Interplay of the mean-field potentials and corrections on threshold energies associated with production cross sections of particles impacts the constraining of stiffness of symmetry energy. The in-medium effects on pions and $\Delta(1232)$ dynamics in heavy-ion collisions have been studied in Refs \cite{Fe05,So15} for threshold energy corrections and in Refs \cite{Xi93,Fu97,Ho14,Fe10} for the pion optical potential.

In this work, the dynamics of pseudoscalar mesons and hyperons with s=-1 in heavy-ion collisions and the in-medium properties of particles in dense nuclear matter are to be investigated with the Lanzhou quantum molecular dynamics (LQMD) transport model. The high-density behavior of nuclear symmetry energy from isospin particles and eta production will be explored. The article is organized as follows. In section II we give a brief description of the recent version of the LQMD model. The in-medium properties and isospin effects on particle dynamics are discussed in section III. Summary and perspective on the mechanism of particle production near threshold energies are presented in section IV.

\section{II. Model description}

In the LQMD model, the dynamics of the resonances ($\Delta$(1232), N*(1440), N*(1535), etc), hyperons ($\Lambda$, $\Sigma$, $\Xi$, $\Omega$) and mesons ($\pi$, $\eta$, $K$, $\overline{K}$, $\rho$, $\omega$) is described via hadron-hadron collisions, decays of resonances, mean-field potentials, and corrections on threshold energies of elementary cross sections \cite{Fe10,Fe09}. Besides the hadron-hadron collisions, we have further included the annihilation channels, charge-exchange reaction, elastic and inelastic collisions in antinucleon-nucleon collisions for understanding antiproton induced reactions \cite{Fe14}.

The temporal evolutions of the baryons (nucleons and resonances) and mesons in the reaction system under the self-consistently generated mean-field are governed by Hamilton's equations of motion, which read as
\begin{eqnarray}
\dot{\mathbf{p}}_{i} = -\frac{\partial H}{\partial\mathbf{r}_{i}}, \quad \dot{\mathbf{r}}_{i} = \frac{\partial H}{\partial\mathbf{p}_{i}}.
\end{eqnarray}
The Hamiltonian of baryons consists of the relativistic energy, the effective interaction potential and the momentum dependent part as follows:
\begin{equation}
H_{B}=\sum_{i}\sqrt{\textbf{p}_{i}^{2}+m_{i}^{2}}+U_{int}+U_{mom}.
\end{equation}
Here the $\textbf{p}_{i}$ and $m_{i}$ represent the momentum and the mass of the baryons.

The effective interaction potential is composed of the Coulomb interaction and the local interaction potential
\begin{equation}
U_{int}=U_{Coul}+U_{loc}.
\end{equation}
The Coulomb interaction potential is written as
\begin{equation}
U_{Coul}=\frac{1}{2}\sum_{i,j,j\neq i} \frac{e_{i}e_{j}}{r_{ij}}erf(r_{ij}/\sqrt{4L})
\end{equation}
where the $e_{j}$ is the charged number including protons and charged resonances. The $r_{ij}=|\mathbf{r}_{i}-\mathbf{r}_{j}|$ is the relative distance of two charged particles, and the $L$ being the square of the Gaussian wave-packet width.

The local interaction potential is derived from the Skyrme energy-density functional as the form of
$U_{loc}=\int V_{loc}(\rho(\mathbf{r}))d\mathbf{r}$. The energy-density functional reads
\begin{eqnarray}
V_{loc}(\rho)=&& \frac{\alpha}{2}\frac{\rho^{2}}{\rho_{0}} +
\frac{\beta}{1+\gamma}\frac{\rho^{1+\gamma}}{\rho_{0}^{\gamma}} + E_{sym}^{loc}(\rho)\rho\delta^{2}
\nonumber \\
&& + \frac{g_{sur}}{2\rho_{0}}(\nabla\rho)^{2} + \frac{g_{sur}^{iso}}{2\rho_{0}}[\nabla(\rho_{n}-\rho_{p})]^{2},
\end{eqnarray}
where the $\rho_{n}$, $\rho_{p}$ and $\rho=\rho_{n}+\rho_{p}$ are the neutron, proton and total densities, respectively, and the $\delta=(\rho_{n}-\rho_{p})/(\rho_{n}+\rho_{p})$ being the isospin asymmetry. The coefficients $\alpha$, $\beta$, $\gamma$, $g_{sur}$, $g_{sur}^{iso}$ and $\rho_{0}$ are set to be the values of -215.7 MeV, 142.4 MeV, 1.322, 23 MeV fm$^{2}$, -2.7 MeV fm$^{2}$ and 0.16 fm$^{-3}$, respectively. A Skyrme-type momentum-dependent potential is used in the LQMD model \cite{Fe12}
\begin{eqnarray}
U_{mom}=&& \frac{1}{2\rho_{0}}\sum_{i,j,j\neq i}\sum_{\tau,\tau'}C_{\tau,\tau'}\delta_{\tau,\tau_{i}}\delta_{\tau',\tau_{j}}\int\int\int d \textbf{p}d\textbf{p}'d\textbf{r}   \nonumber \\
&& \times f_{i}(\textbf{r},\textbf{p},t) [\ln(\epsilon(\textbf{p}-\textbf{p}')^{2}+1)]^{2} f_{j}(\textbf{r},\textbf{p}',t).
\end{eqnarray}
Here $C_{\tau,\tau}=C_{mom}(1+x)$, $C_{\tau,\tau'}=C_{mom}(1-x)$ ($\tau\neq\tau'$) and the isospin symbols $\tau$($\tau'$) represent proton or neutron. The parameters $C_{mom}$ and $\epsilon$ was determined by fitting the real part of optical potential as a function of incident energy from the proton-nucleus elastic scattering data. In the calculation, we take the values of 1.76 MeV, 500 c$^{2}$/GeV$^{2}$ for the $C_{mom}$ and $\epsilon$, respectively, which result in the effective mass $m^{\ast}/m$=0.75 in nuclear medium at saturation density for symmetric nuclear matter. The parameter $x$ as the strength of the isospin splitting with the value of -0.65 is taken in this work, which has the mass splitting of $m^{\ast}_{n}>m^{\ast}_{p}$ in nuclear medium.  A compression modulus of K=230 MeV for isospin symmetric nuclear matter is concluded in the LQMD model.

The symmetry energy is composed of three parts, namely the kinetic energy of free Fermi gas, the local density-dependent interaction and the momentum-dependent potential as
\begin{equation}
E_{sym}(\rho)=\frac{1}{3}\frac{\hbar^{2}}{2m}\left(\frac{3}{2}\pi^{2}\rho\right)^{2/3}+E_{sym}^{loc}(\rho)+E_{sym}^{mom}(\rho).
\end{equation}
The local part is adjusted to mimic predictions of the symmetry energy calculated by microscopical or phenomenological many-body theories and has two-type forms as follows:
\begin{equation}
E_{sym}^{loc}(\rho)=\frac{1}{2}C_{sym}(\rho/\rho_{0})^{\gamma_{s}},
\end{equation}
and
\begin{equation}
E_{sym}^{loc}(\rho)=a_{sym}(\rho/\rho_{0})+b_{sym}(\rho/\rho_{0})^{2}.
\end{equation}
The parameters $C_{sym}$, $a_{sym}$ and $b_{sym}$ are taken as the values of 52.5 MeV, 43 MeV, -16.75 MeV. The values of $\gamma_{s}$=0.5, 1, 2 lead to the soft, linear and hard symmetry energy in the domain of high densities, respectively, and the Eq. (9) gives a supersoft symmetry energy, which cover the largely uncertain of nuclear symmetry energy, particularly at supra-saturation densities. All cases cross at saturation density with the value of 31.5 MeV. The values of slope parameters $L=3\rho_{0}(\partial E_{sym}/\partial \rho)|_{\rho=\rho_{0}}$ are 203.7 MeV, 124.9 MeV, 85.6 MeV and 74.7 MeV for the hard, linear, soft and supersoft symmetry energies, respectively. And the corresponding 448 MeV, -24.5 MeV, -83.5 MeV and -326 MeV for the curvature parameters $K_{sym}=9\rho_{0}^{2}(\partial^{2} E_{sym}/\partial \rho^{2})|_{\rho=\rho_{0}}$ are concluded. It should be mentioned the short-range correlation of Fermi gas reduces the kinetic energy part of symmetry energy \cite{He15}.

The hyperon mean-field potential is constructed on the basis of the light-quark counting rule. The self-energies of hyperons are assumed to be two thirds of that experienced by nucleons. Thus, the in-medium dispersion relation reads
\begin{equation}
\omega(\textbf{p}_{i},\rho_{i})=\sqrt{(m_{H}+\Sigma_{S}^{H})^{2}+\textbf{p}_{i}^{2}} + \Sigma_{V}^{H}
\end{equation}
with $\Sigma_{S}^{H}= 2 \Sigma_{S}^{N}/3$ and $\Sigma_{V}^{H}= 2 \Sigma_{V}^{N}/3$, which leads to the optical potential at the saturation density being the value of -32 MeV. The antinucleon-nucleon potential is similar to hyperons,
A factor $\xi$ is introduced to mimic the antiproton-nucleus scattering \cite{La09} and the real part of phenomenological antinucleon-nucleon optical potential \cite{Co82} as $\Sigma_{S}^{\overline{N}}=\xi\Sigma_{S}^{N}$ and $\Sigma_{V}^{\overline{N}}=-\xi\Sigma_{V}^{N}$ with $\xi$=0.25, which leads to the optical potential $V_{\overline{N}}$=-164 MeV for an antinucleon at the zero momentum at and normal nuclear density $\rho_{0}$=0.16 fm$^{-3}$.

The Hamiltonian of mesons (here mainly concentrating on pseudoscalar mesons) is constructed as follows
\begin{eqnarray}
H_{M} = \sum_{i=1}^{N_{M}}\left( V_{i}^{\textrm{Coul}} + \omega(\textbf{p}_{i},\rho_{i}) \right).
\end{eqnarray}
Here the Coulomb interaction is given by
\begin{equation}
V_{i}^{\textrm{Coul}}=\sum_{j=1}^{N_{B}}\frac{e_{i}e_{j}}{r_{ij}},
\end{equation}
where the $N_{M}$ and $N_{B}$ are the total numbers of mesons and baryons including charged resonances, respectively. The energy of pion in the nuclear medium is composed of the isoscalar and isovector contributions as follows
\begin{equation}
\omega_{\pi}(\textbf{p}_{i},\rho_{i}) = \omega_{isoscalar}(\textbf{p}_{i},\rho_{i})+C_{\pi}\tau_{z}\delta (\rho/\rho_{0})^{\gamma_{\pi}}.
\end{equation}
The coefficient $C_{\pi}= \rho_{0} \hbar^{3}/(4f^{2}_{\pi}) = 36$ MeV, and the isospin quantity $\tau_{z}=$ 1, 0, and -1 for $\pi^{-}$, $\pi^{0}$ and $\pi^{+}$, respectively. The isospin asymmetry $\delta=(\rho_{n}-\rho_{p})/(\rho_{n}+\rho_{p})$ and the quantity $\gamma_{\pi}$ adjusts the isospin splitting of pion optical potential. Usually, we take the $\gamma_{p} = 2$ in the model. Here, we have two choices in evaluation of the isoscalar part, the phenomenological ansatz \cite{Ga87} and the $\Delta$-hole model \cite{Br75}. With the framework of the phenomenological ansatz, the dispersion relation reads
\begin{equation}
\omega_{isoscalar}(\textbf{p}_{i},\rho_{i})=\sqrt{(|\textbf{p}_{i}|-p_{0})^{2}+m_{0}^{2}}-U,
\end{equation}
with
\begin{eqnarray}
U=\sqrt{p_{0}^{2}+m_{0}^{2}}-m_{\pi},  \\
m_{0}=m_{\pi}+6.5(1-x^{10})m_{\pi},   \\
p_{0}^{2}=(1-x)^{2}m_{\pi}^{2}+2m_{0}m_{\pi}(1-x).
\end{eqnarray}
The phenomenological medium dependence on the baryon density is introduced via the coefficient $x(\rho_{i})=\exp(-a(\rho_{i}/\rho_{0}))$ with the parameter $a=0.154$ and the saturation density $\rho_{0}$ in nuclear matter. Impact of the phenomenological approach on pion dynamics in heavy-ion collisions has been investigated in Refs \cite{Fe10,Fu97}.

\begin{figure*}
\includegraphics{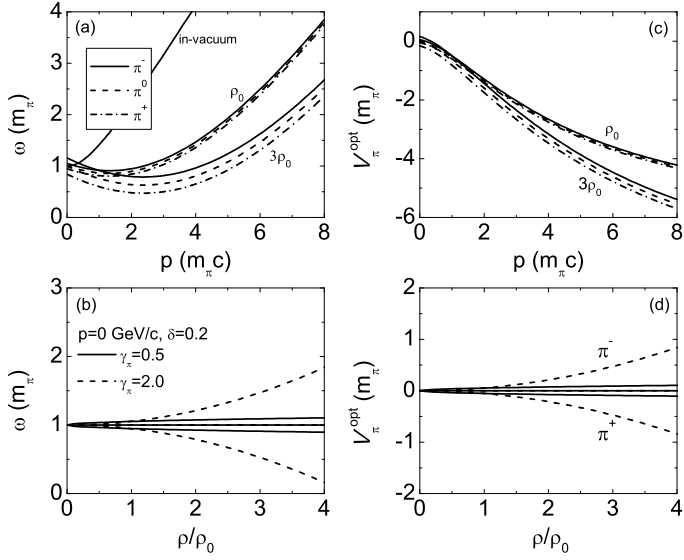}
\caption{\label{fig:wide} Momentum and density dependence of the pion energy and the optical potential in dense nuclear matter.}
\end{figure*}

On the other hand, the pion self-energy in the nuclear medium have been studied via the $\Delta$-hole model \cite{Br75}. The in-medium pion dispersion relation consists of a pion branch (smaller value) and a $\Delta$-hole (larger value) branch, which become softened and hardened with baryon density in nuclear matter, respectively. Thus, the dispersion relation reads
\begin{eqnarray}
\omega_{isoscalar}(\textbf{p}_{i},\rho_{i}) = &&  S_{\pi}(\textbf{p}_{i},\rho_{i}) \omega_{\pi-like}(\textbf{p}_{i},\rho_{i}) +    \nonumber \\
&&        S_{\Delta}(\textbf{p}_{i},\rho_{i}) \omega_{\Delta-like}(\textbf{p}_{i},\rho_{i}) .
\end{eqnarray}
The probabilities of the pion component satisfy the relation
\begin{equation}
S_{\pi}(\textbf{p}_{i},\rho_{i}) + S_{\Delta}(\textbf{p}_{i},\rho_{i}) = 1
\end{equation}
The value of the probability is determined from the pion self-energy as \cite{Xi93}
\begin{equation}
S(\textbf{p}_{i},\rho_{i}) = \frac{1}{1-\partial \Pi (\omega)/\partial\omega^{2}},
\end{equation}
where the pion self-energy is given by
\begin{equation}
\Pi = \textbf{p}_{i}^{2}\frac{\chi}{1 - g\prime\chi},
\end{equation}
with the Migdal parameter $g\prime\sim$0.6 and
\begin{equation}
\chi = -\frac{8}{9}\left(\frac{f_{\Delta}}{m_{\pi}}\right)^{2} \frac{\omega_{\Delta}\rho\hbar^{3}}{\omega_{\Delta}^{2}-\omega^{2}} \exp(-2\textbf{p}_{i}^{2}/b^{2}).
\end{equation}
The $\omega_{\Delta}=\sqrt{m_{\Delta}^{2}+\textbf{p}_{i}^{2}}-m_{N}$, the $m_{\pi}$, $m_{N}$ and $m_{\Delta}$ are the masses of pion, nucleon and delta, respectively. The $\pi N\Delta$ coupling constant $f_{\Delta}\sim 2 $ and the cutoff factor $b\sim 7 m_{\pi}$. Two eigenvalues of $\omega_{\pi-like}$ and $\omega_{\Delta-like}$ are obtained from the pion dispersion relation as
\begin{equation}
\omega^{2} = \textbf{p}_{i}^{2} + m_{\pi}^{2} + \Pi(\omega).
\end{equation}
The energy balance in the decay of resonances is satisfied with the relation
\begin{equation}
\sqrt{m_{R}^{2}+\textbf{p}_{R}^{2}} = \sqrt{m_{N}^{2}+(\textbf{p}_{R} -\textbf{p}_{\pi})^{2}} + \omega_{\pi}(\textbf{p}_{\pi},\rho) + V_{\pi}^{\textrm{Coul}},
\end{equation}
where the $\textbf{p}_{R}$ and $\textbf{p}_{\pi}$ are the momenta of resonances and pions, respectively.
The optical potential can be evaluated from the in-medium energy $V_{\pi}^{opt}(\textbf{p}_{i},\rho_{i}) = \omega_{\pi}(\textbf{p}_{i},\rho_{i}) - \sqrt{m_{\pi}^{2}+\textbf{p}_{i}^{2}}$. Shown in Fig. 1 is the pion energy and optical potential as functions of the pion momentum and baryon density in units of pion mass and saturation density. The attractive potential is obtained with increasing the pion momentum. Isospin splitting of the pion potential appears and the effect is pronounced in the domain of high baryon density, which impacts the charged pion ratios in heavy-ion collisions. Influence of the in-medium effects on the charged pion ratio is also investigated within a thermal model in Ref. \cite{Xu10}.

\begin{figure*}
\includegraphics{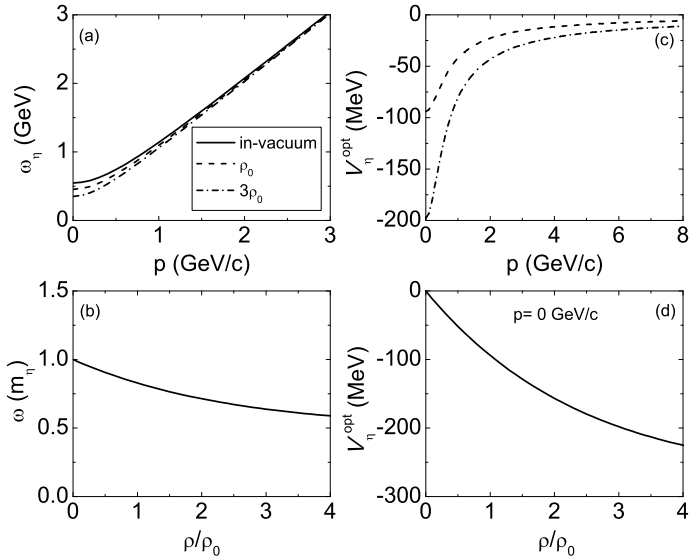}
\caption{\label{fig:wide} The energy and optical potential as functions of  $\eta$ momentum and baryon density.}
\end{figure*}

The Hamiltonian of $\eta$ is composed of
\begin{equation}
H_{\eta} = \sum_{i=1}^{N_{\eta}}\left( \sqrt{m_{\eta}^{2}+\textbf{p}_{i}^{2}} + V_{\eta}^{opt}(\textbf{p}_{i},\rho_{i}) \right).
\end{equation}
The eta optical potential is evaluated from the dispersion relation based on the chiral perturbation theory \cite{Ni08} as
\begin{equation}
\omega_{\eta}(\textbf{p}_{i},\rho_{i}) = \sqrt{\left(m^{2}_{\eta} - a_{\eta}\rho_{s}\right) \left(1+b_{\eta}\rho_{s}\right)^{-1} + \textbf{p}_{i}^{2}}
\end{equation}
with $a_{\eta}=\hbar^{3}\frac{\Sigma_{\eta N}}{f^{2}_{\pi}}$ and $b_{\eta}=\hbar^{2}\frac{\kappa}{f^{2}_{\pi}}$. The pion decay constant $f_{\pi}=$92.4 MeV, $\Sigma_{\eta N}=$280 MeV and $\kappa =$0.4 fm. The optical potential is given by $V_{\eta}^{opt}(\textbf{p}_{i},\rho_{i}) = \omega_{\eta}(\textbf{p}_{i},\rho_{i}) - \sqrt{m_{\eta}^{2}+\textbf{p}_{i}^{2}}$ with the eta mass $m_{\eta}$=547 MeV. The value of $V_{\eta}^{opt}=$ -94 MeV is obtained with zero momentum and saturation density $\rho=\rho_{0}$. The attractive potential is used in this work as shown in Fig. 2. Up to now, although it has not consistent conclusions on the depth of the $\eta$ potential in nucleus. Basically the attractive $\eta$-nucleon interaction got to be favored by different models \cite{Ch91}. Even the existence of bound $\eta$-nucleus was pointed out \cite{Li86}.

\begin{figure*}
\includegraphics{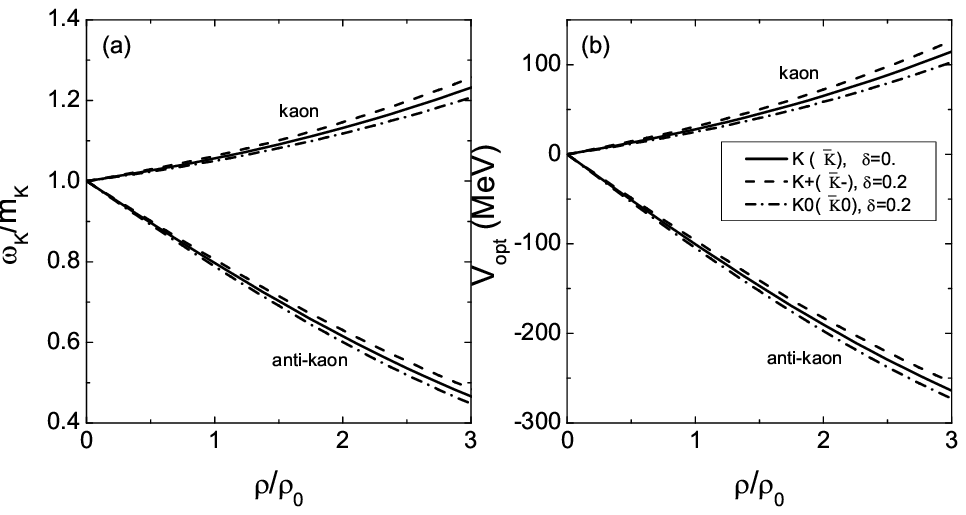}
\caption{\label{fig:wide} Density dependence of the kaon energy and the optical potential in isospin symmetric and asymmetric nuclear matter ($\delta$=0.2).}
\end{figure*}

The kaon and anti-kaon energies in the nuclear medium distinguish isospin effects based on the chiral Lagrangian approach as \cite{Ka86,Sc97,Fe13}
\begin{eqnarray}
\omega_{K}(\textbf{p}_{i},\rho_{i})= && \left[m_{K}^{2}+\textbf{p}_{i}^{2}-a_{K}\rho_{i}^{S}
-\tau_{3}c_{K}\rho_{i3}^{S}+(b_{K}\rho_{i}+\tau_{3}d_{K}\rho_{i3})^{2}\right]^{1/2}
\nonumber \\
&& +b_{K}\rho_{i}+\tau_{3}d_{K}\rho_{i3}
\end{eqnarray}
and
\begin{eqnarray}
\omega_{\overline{K}}(\textbf{p}_{i},\rho_{i})= && \left[m_{\overline{K}}^{2}+\textbf{p}_{i}^{2}-a_{\overline{K}}\rho_{i}^{S}
-\tau_{3}c_{K}\rho_{i3}^{S}+(b_{K}\rho_{i}+\tau_{3}d_{K}\rho_{i3})^{2}\right]^{1/2}
\nonumber \\
&& -b_{K}\rho_{i}-\tau_{3}d_{K}\rho_{i3},
\end{eqnarray}
respectively. Here the $b_{K}=3/(8f_{\pi}^{\ast 2})\approx$0.333 GeVfm$^{3}$, the $a_{K}$ and $a_{\overline{K}}$ are 0.18 GeV$^{2}$fm$^{3}$ and 0.31 GeV$^{2}$fm$^{3}$, respectively, which result in the strengths of repulsive kaon-nucleon (KN) potential and of attractive antikaon-nucleon $\overline{K}$N potential with the values of 27.8 MeV and -100.3 MeV at saturation baryon density for isospin symmetric matter, respectively. The $\tau_{3}$=1 and -1 for the isospin pair K$^{+}$($\overline{K}^{0}$) and K$^{0}$(K$^{-}$), respectively. The parameters $c_{K}$=0.0298 GeV$^{2}$fm$^{3}$ and $d_{K}$=0.111 GeVfm$^{3}$ determine the isospin splitting of kaons in neutron-rich nuclear matter. The optical potential of kaon is derived from the in-medium energy as $V_{opt}(\textbf{p},\rho)=\omega(\textbf{p},\rho)-\sqrt{\textbf{p}^{2}+m_{K}^{2}}$. The values of $m^{\ast}_{K}/m_{K}$=1.056 and $m^{\ast}_{\overline{K}}/m_{\overline{K}}$=0.797 at normal baryon density are concluded with the parameters in isospin symmetric nuclear matter. The effective mass $m^{\ast}=\omega(\textbf{p}=0,\rho=\rho_{0})$ is used to evaluate the threshold energy for kaon and antikaon production, e.g., the threshold energy in the pion-baryon collisions $\sqrt{s_{th}}=m^{\ast}_{Y} + m^{\ast}_{K}$. Shown in Fig. 3 is a comparison of the kaon energy in units of its free mass and the optical potential at the momentum of $\textbf{p}$=0 as a function of baryon density. In the neutron-rich nuclear matter, an isospin splitting for the pairs (K$^{0}$, K$^{+}$) and ($\overline{K}^{0}$, K$^{-}$) is pronounced with the baryon density. The equations of motion include the spatial component of the vector potential, which would lead to an attractive Lorentz force between kaons and nucleons as \cite{Zh04}
\begin{equation}
\frac{d\mathbf{p}_{i}}{dt}=-\frac{\partial V_{i}^{\textrm{Coul}}}{\partial\mathbf{r}_{i}}-
\frac{\partial\omega_{K(\overline{K})}(\textbf{p}_{i},\rho_{i})}{\partial\mathbf{r}_{i}}\pm
\textbf{v}_{i}\frac{\partial \textbf{V}_{i}}{\partial\mathbf{r}_{i}},
\end{equation}
where $\textbf{v}_{i}$ is the velocity of kaons (anti-kaons), and $\pm$ for K and $\overline{K}$, respectively. The Lorentz force increases with the kaon momentum. However this contradicts the predictions from the theoretical studies which show an opposite momentum dependence \cite{Sh92}. The isospin effect was considered in the vector potential $\textbf{V}_{i}$ \cite{Sc97,Fe13}. The KN potential and the Lorentz force influence the kaon dynamics in heavy-ion collisions and in proton induced reactions, which lead to a flat structure of direct flows and well reproduce the experimental data \cite{Fe13}.

The scattering in two-particle collisions is performed by using a Monte Carlo procedure, in which the probability to be a channel in a collision is calculated by its contribution of the channel cross section to the total cross section. The primary products in nucleon-nucleon (NN) collisions are the resonances of $\Delta$(1232), $N^{\ast}$(1440), and $N^{\ast}$(1535). We have included the reaction channels as follows:
\begin{eqnarray}
&& NN \leftrightarrow N\triangle, \quad  NN \leftrightarrow NN^{\ast}, \quad  NN
\leftrightarrow \triangle\triangle,  \nonumber \\
&& \Delta \leftrightarrow N\pi,  N^{\ast} \leftrightarrow N\pi,  NN \leftrightarrow NN\pi (s-state),  \nonumber \\
&& N^{\ast}(1535) \leftrightarrow N\eta.
\end{eqnarray}
Here hadron-hadron collisions take place as two-body process and three-body ($s-$state pion production) reactions.
At the considered energies, there are mostly $\Delta$ resonances which disintegrate into a $\pi$ and a nucleon in the evolutions. The momentum-dependent decay widths are used for the resonances of $\Delta$(1232) and $N^{\ast}$(1440) \cite{Fe09,Hu94}. We have taken a constant width of $\Gamma$=150 MeV for the $N^{\ast}$(1535) decay. Elastic scattering of NN, nucleon-resonance ($NR\rightarrow NR$) and resonance-resonance ($RR\rightarrow RR$) collisions and inelastic collisions of nucleon-resonance ($NR\rightarrow NN$, $NR\rightarrow NR\prime$) and resonance-resonance ($RR\rightarrow NN$, $RR\rightarrow RR\prime$, $R$ and $R\prime$ being different resonances), have been included in the model.

The elastic cross sections of the available experimental data are parameterized in the energy range of 1 MeV - 2 TeV for NN collisions \cite{Fe09,Fe12b}, in which the $pn$ (proton-neutron) cross sections are about 3 times larger than the $pp$ (proton-proton)/$nn$ (neutron-neutron) cases at the incident energies from 1 MeV to 400 MeV. The in-medium elastic cross section is scaled according to the effective mass through $\sigma_{NN}^{medium}=(\mu^{\ast}_{NN}/\mu_{NN})^{2}\sigma_{NN}^{free}$ with the $\mu^{\ast}_{NN}$ and $\mu_{NN}$ being the reduced masses of colliding nucleon pairs in the medium and in the free space, respectively \cite{Fe12b}. The parameterized cross sections calculated by the one-boson exchange model \cite{Hu94} for the $\Delta$(1232) and $N^{\ast}$(1440) production are used in the model. The cross sections for $N^{\ast}$(1535) are estimated from the empirical $\eta$ production, such as $\sigma(pp(nn) \rightarrow NN^{\ast}(1535)) \approx 2\sigma(pp(nn) \rightarrow pp(nn)\eta)=$ (a) $0.34 s_{r}/(0.253+s_{r}^{2})$, (b) $0.4 s_{r}/(0.552+s_{r}^{2})$ and (c) $0.204 s_{r}/(0.058+s_{r}^{2})$ in mb and $s_{r}=\sqrt{s}-\sqrt{s_{0}}$ with $\sqrt{s}$ being the invariant energy in GeV and $\sqrt{s_{0}}=2m_{N}+m_{\eta}=2.424$ GeV. The case (b) is used in this work. The $np$ cross sections are about 3 times larger than that for $nn$. Half probabilities of the resonances $N^{\ast}(1535)$ decay into the $\eta$ production. Different stiffness of symmetry energy leads to difference of $nn (pp)$ and $np$ collision probabilities in producing $N^{\ast}(1535)$. Therefore, the production of $\eta$ could be probe of the high-density symmetry energy.

The strangeness and vector mesons ($\rho$, $\omega$) are created in inelastic hadron-hadron collisions without intermediate resonances. We included the channels as follows:
\begin{eqnarray}
&& BB \rightarrow BYK,  BB \rightarrow BBK\overline{K},  B\pi(\eta) \rightarrow YK,  YK \rightarrow B\pi,     \nonumber \\
&& B\pi \rightarrow NK\overline{K}, Y\pi \rightarrow B\overline{K}, \quad  B\overline{K} \rightarrow Y\pi, \quad YN \rightarrow \overline{K}NN,  \nonumber \\
&& NN \rightarrow NN\rho, NN \rightarrow NN\omega.
\end{eqnarray}
Here the B stands for (N, $\triangle$, N$^{\ast}$) and Y($\Lambda$, $\Sigma$), K(K$^{0}$, K$^{+}$) and $\overline{K}$($\overline{K}^{0}$, K$^{-}$). The parameterized cross sections of each isospin channel $BB \rightarrow BYK$ \cite{Ts99} are used in the calculation. We take the parametrizations of the channels $B\pi \rightarrow YK$ \cite{Ts94} besides the $N\pi \rightarrow \Lambda K$ reaction \cite{Cu84}. The results are close to the experimental data at near threshold energies. The cross section of antikaon production in inelastic hadron-hadron collisions is taken as the same form of the parametrization used in the hadron string dynamics (HSD) calculations \cite{Ca97}. Furthermore, the elastic scattering and strangeness-exchange reaction between strangeness and baryons have been considered through the channels of $KB \rightarrow KB$, $YB \rightarrow YB$ and $\overline{K}B \rightarrow \overline{K}B$ and we use the parametrizations in Ref. \cite{Cu90}. The charge-exchange reactions between the $KN \rightarrow KN$ and $YN \rightarrow YN$ channels are included by using the same cross sections with the elastic scattering, such as $K^{0}p\rightarrow K^{+}n$, $K^{+}n\rightarrow K^{0}p$ etc \cite{Fe13}. The cross sections for $\rho$ and $\omega$ production are taken from the fitting of experimental data \cite{Li01}.

\section{III. Results and discussions}

Superdense hadronic matter can be formed in high-energy heavy-ion collisions and exists in the compact stars, such as neutron stars. The hadron-hadron interaction in the superdense matter is complicated and varying with the baryon density. The hadron-nucleon potential impacts the ingredients of hadrons in the compact stars. In nuclear reactions, i.e., heavy-ion collisions, antiproton (proton) induced reactions etc, the production and phase-space distribution of particles were modified in nuclear medium \cite{Fe13}. The isospin dependence of the potential and the corrections on threshold energies influence the ratios of isospin particles. Consequently, the extraction of high-density symmetry energy is to be moved from particle production. On the other hand, the yields of particles and bound fragments such as hypernuclides, kaonic nucleus, antiprotonic nucleus is related to the potential of particles in nuclear medium.

\begin{figure*}
\includegraphics{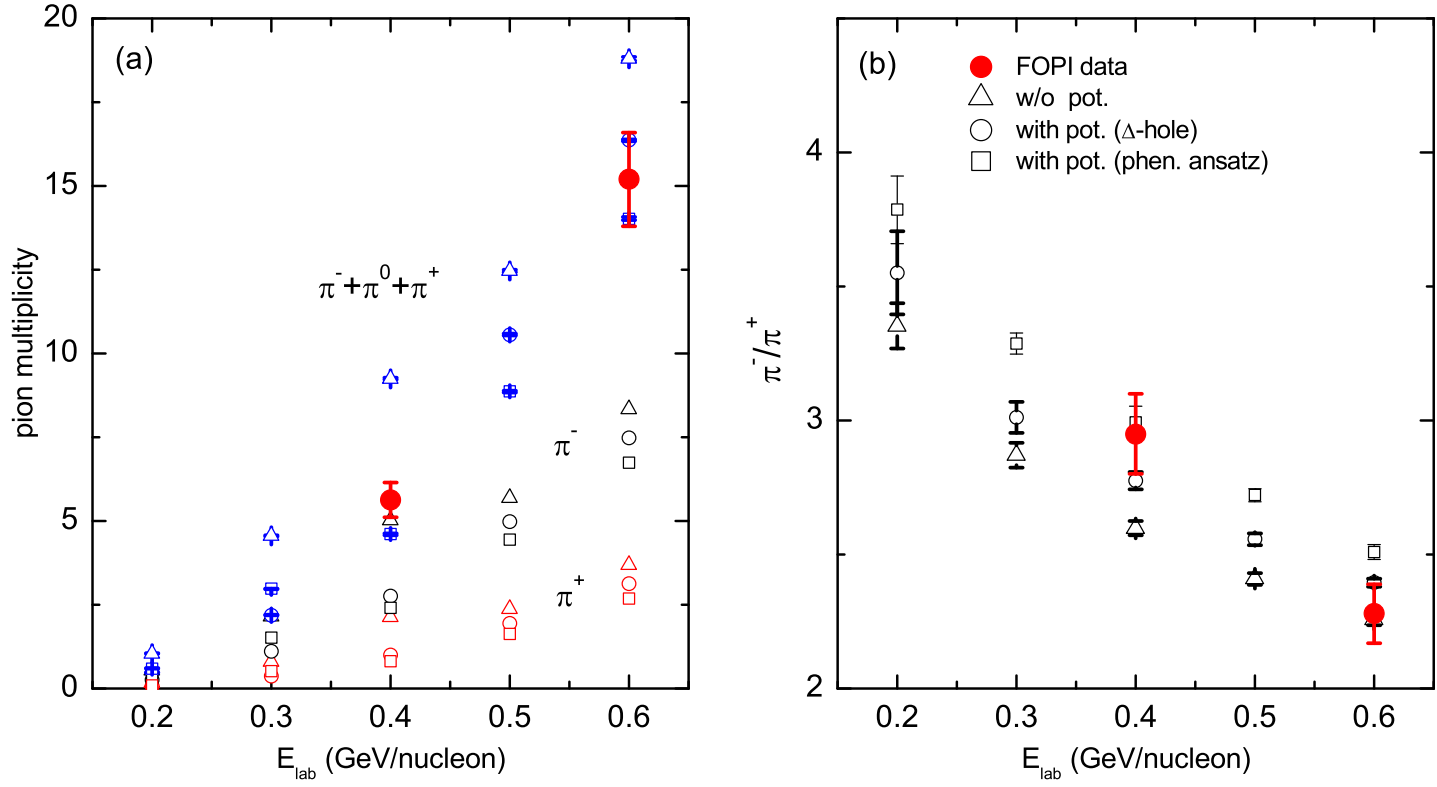}
\caption{\label{fig:wide} Total multiplicity of pion and ratio of charged pion produced in central $^{197}$Au+$^{197}$Au collisions. The data from FOPI collaboration \cite{Re07} is compared to different cases of pion-nucleon potentials.}
\end{figure*}

\begin{figure*}
\includegraphics{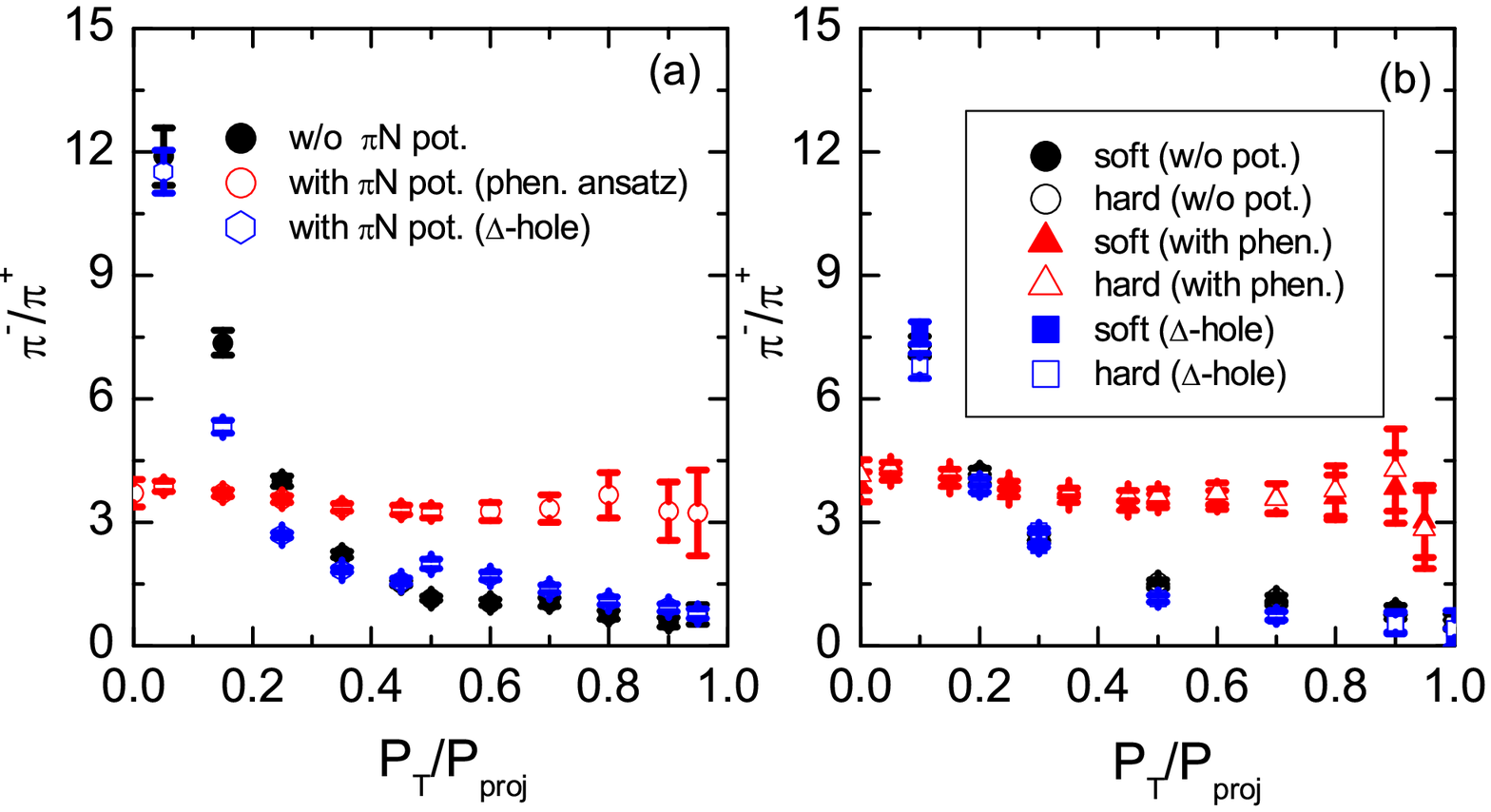}
\caption{\label{fig:wide} Transverse momentum distribution of the $\pi^{-}/\pi^{+}$ ratio in central $^{197}$Au+$^{197}$Au collisions at incident energy of 300 MeV/nucleon.}
\end{figure*}

Pion as the lightest meson could be easily produced in heavy-ion collisions and attracts much attention both in theories and in experiments. Besides as probing the high-density symmetry energy, the pion-nucleon potential in dense matter is not well understood up to now. In this work, we thoroughly investigated the pion dynamics in heavy-ion collisions, in particular in the domain of near threshold energies (E$_{th}$=280 MeV for $\pi^{0}$). Impacts of the isospin and momentum dependent pion potential and the stiffness of symmetry energy on the pion production are stressed. Shown in Fig. 4 is a comparison of total pion and ratio of charged pion produced in central $^{197}$Au+$^{197}$Au collisions with the experimental data from FOPI collaboration \cite{Re07}. It is obvious that the total number of pions is reduced with inclusion of the pion-nucleon potential. However, the $\pi^{-}/\pi^{+}$ ratio is enhanced, in particular with the phenomenological approach for the isoscalar part of the pion energy in medium. The decrease of the pion yields is caused from the attractive interaction between pions and nucleons, which enhances the absorption collisions of pions and nucleons via the channels $\pi N\rightarrow \Delta(1232)$ and $\Delta(1232)N\rightarrow NN$. The effect increases with the pion momentum and baryon density. The isospin related pion-nucleon potential, i.e., more attractive interaction for the $\pi^{+}$N potential, leads to the increase of the $\pi^{-}/\pi^{+}$ ratio. Precise structure of the charged pion ratio can be observed from the transverse momentum distribution as shown in Fig. 5. Similar trends for free pion transportation (only including Coulomb interaction) and the $\Delta$-hole model are found. A flat structure is concluded with the phenomenological approach. However, the effect of symmetry energy from the transverse momentum spectra is negligible for the different $\pi N$ potentials.

To eliminate the Coulomb interaction of charged particles, neutral particles produced in heavy-ion collisions could be nice probes in extracting the in-medium potential. Usually, the neutral particles are reconstructed via the decays in experimentally, e.g., $\pi^{0}\rightarrow 2\gamma$. Shown in Fig. 6 is the $\pi^{0}$ production in $^{197}$Au+$^{197}$Au collisions at the incident energy of 300 MeV/nucleon with different $\pi$N potentials, in which only the isoscalar part of the $\pi$N interaction contributes the pion dynamics. It is obvious that the number of $\pi^{0}$ is reduced from the rapidity and transverse momentum distributions with the $\pi$N potentials. The impact of the $\eta$N potential on $\eta$ dynamics is shown in Fig. 7. One notices that the production cross sections are reduced in the domain of mid-rapidities and high momenta because of attractive interaction between etas and nucleons in nuclear medium. The attractive optical potential enhances the absorption reactions of particles in nuclear medium, in particular at high baryon densities, i.e., via the channels $\pi N\rightarrow \Delta(1232)$, $\eta N\rightarrow N\ast(1535)$. Shown in Fig. 8 is the phase-space distribution of $\eta$ production in central $^{197}$Au+$^{197}$Au collisions at incident energy of 600 MeV/nucleon with different stiffness of symmetry energy. The effect of the $\eta$N potential is similar to the $^{40}$Ca+$^{40}$Ca reaction shown in Fig. 7. However, the difference of hard and soft symmetry energies at high-baryon densities is negligible in the spectra. It is concluded that constraining the symmetry energy from the transverse momentum or kinetic energy spectrum is not possible, but more sensitive to the $\eta$N potential.

\begin{figure*}
\includegraphics{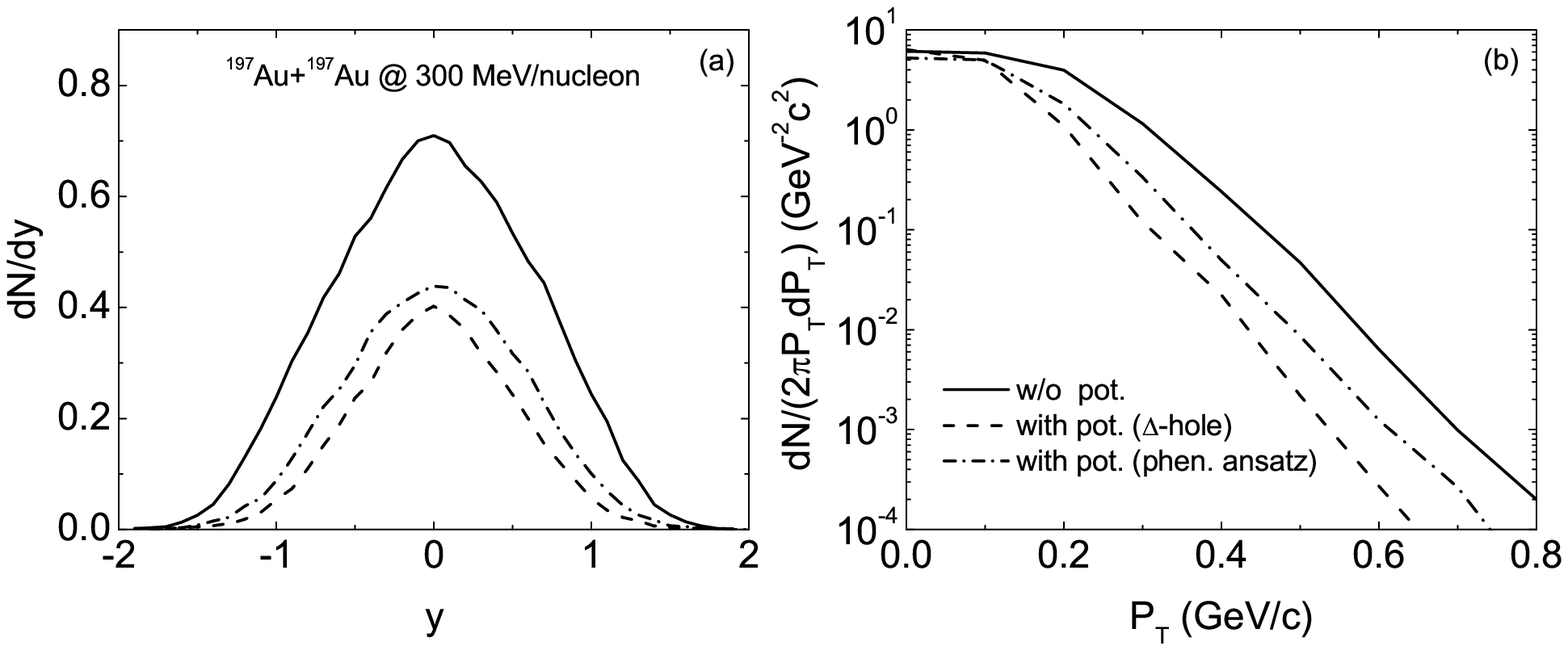}
\caption{\label{fig:wide} Distributions of rapidity and transverse momentum for $\pi^{0}$ production in $^{197}$Au+$^{197}$Au collisions at incident energy of 300 MeV/nucleon.}
\end{figure*}

\begin{figure*}
\includegraphics{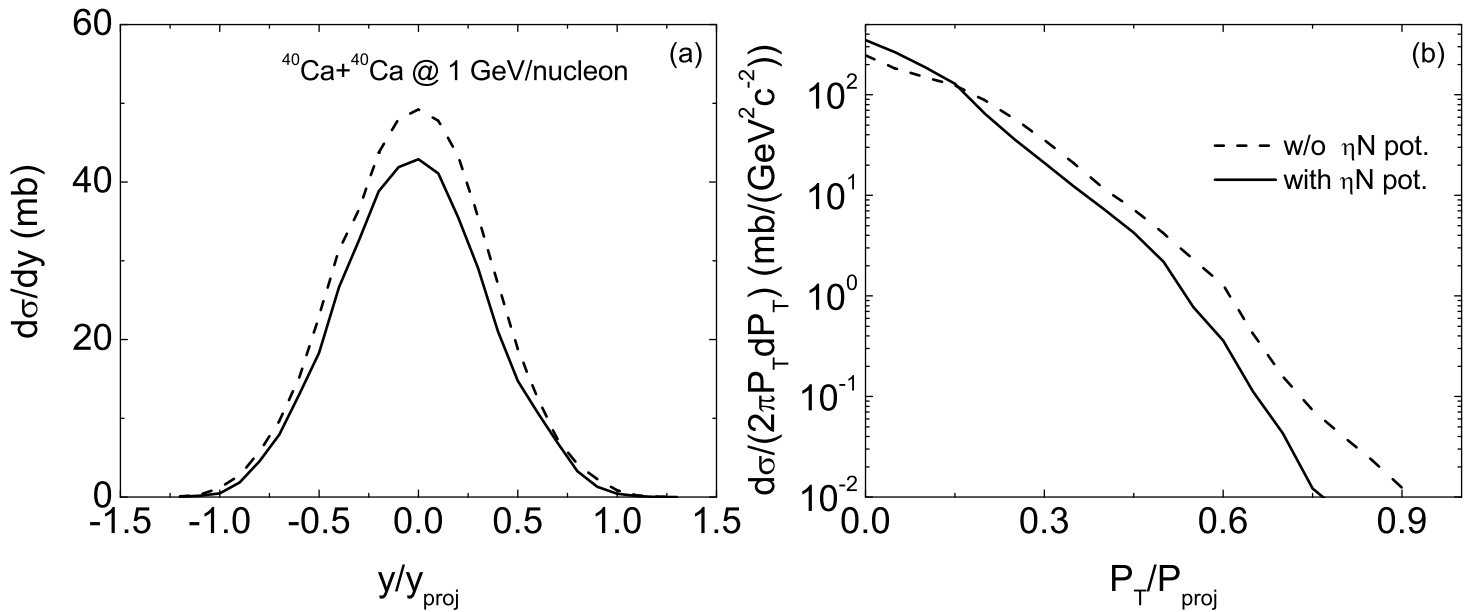}
\caption{\label{fig:wide} Rapidity and transverse momentum spectra of $\eta$ produced in $^{40}$Ca+$^{40}$Ca collisions at incident energy of 1 $\emph{A}$ GeV.}
\end{figure*}

\begin{figure*}
\includegraphics{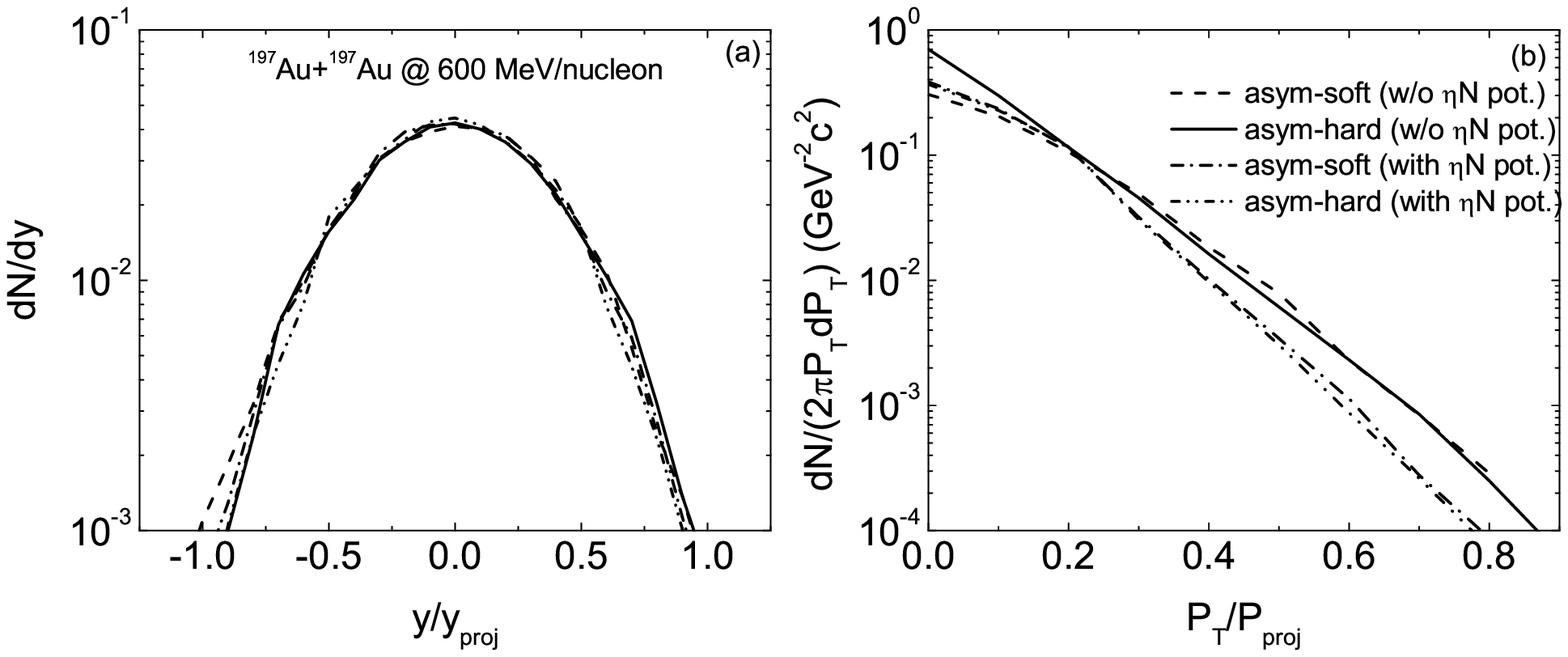}
\caption{\label{fig:wide} Impact of symmetry energy on the $\eta$ production in central $^{197}$Au+$^{197}$Au collisions at incident energy of 600 MeV/nucleon.}
\end{figure*}

A systematic comparison of the in-medium corrections on neutral particle distributions in phase space is shown in Fig. 9. The mean-field potentials of particles in nuclear medium contribute the dynamical evolutions. Consequently, the KN potential reduces the kaon production at midrapidities and at high transverse momenta. However, an opposite contribution of the $\eta$N potential is obtained because of the attractive interaction of eta and nucleon in nuclear medium. The hyperon-nucleon interaction is negligible for $\Lambda$ dynamics. Shown in Fig. 10 is the inclusive spectra of $\pi^{0}$, $\eta$, $K^{0}$, $\overline{K}^{0}$ and neutral hyperons ($\Lambda$+$\Sigma^{0}$) in $^{197}$Au+$^{197}$Au collisions. Similar to the transverse momentum spectra, the eta-nucleon, kaon-nucleon and hyperon-nucleon potentials weakly impact the particle emission. However, the interaction of pions (antikaons) and nucleons contributes the pion (antikaon) dynamics, i.e., increasing the antikaon production at low kinetic energies.

\begin{figure*}
\includegraphics{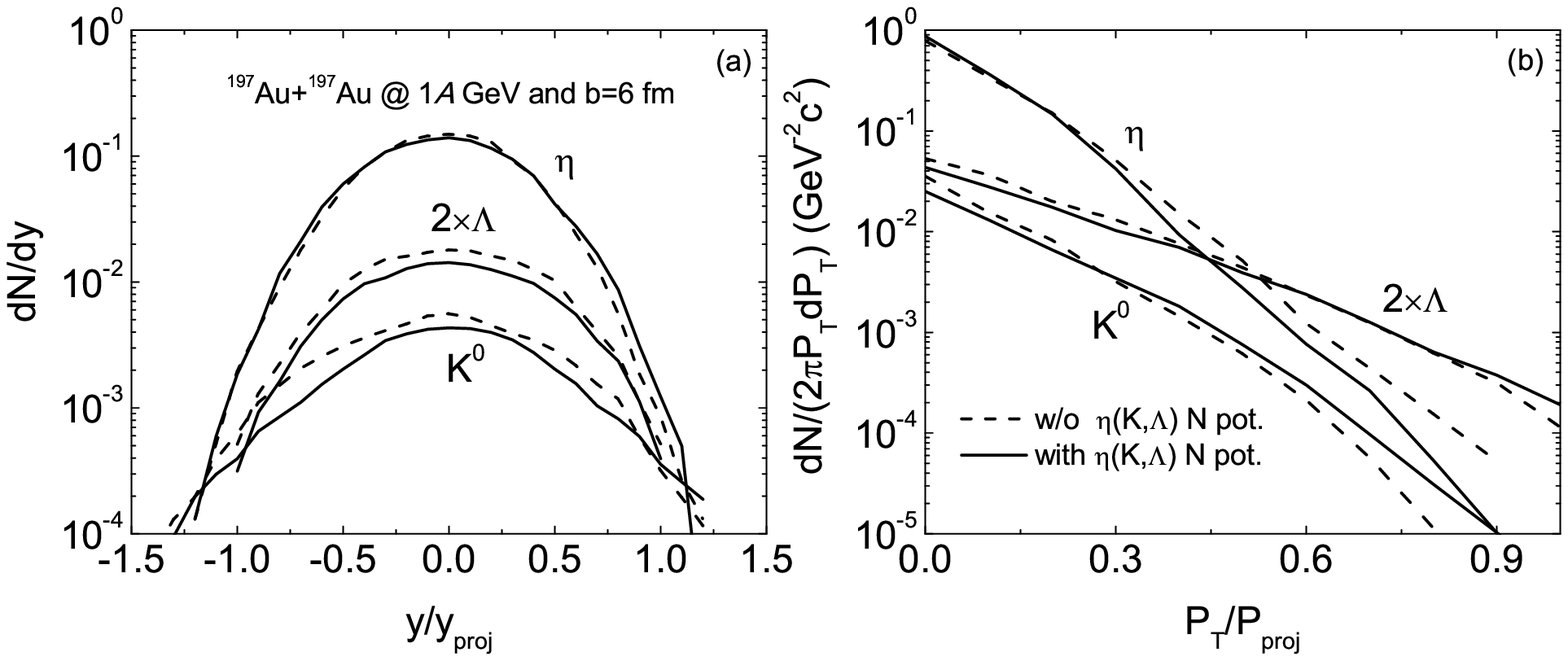}
\caption{\label{fig:wide} Neutral particles $\eta$, K$^{0}$ and $\Lambda$ produced in $^{197}$Au+$^{197}$Au collisions at incident energy of 1\emph{A} GeV and with impact parameter b=6 fm.}
\end{figure*}

\begin{figure*}
\includegraphics{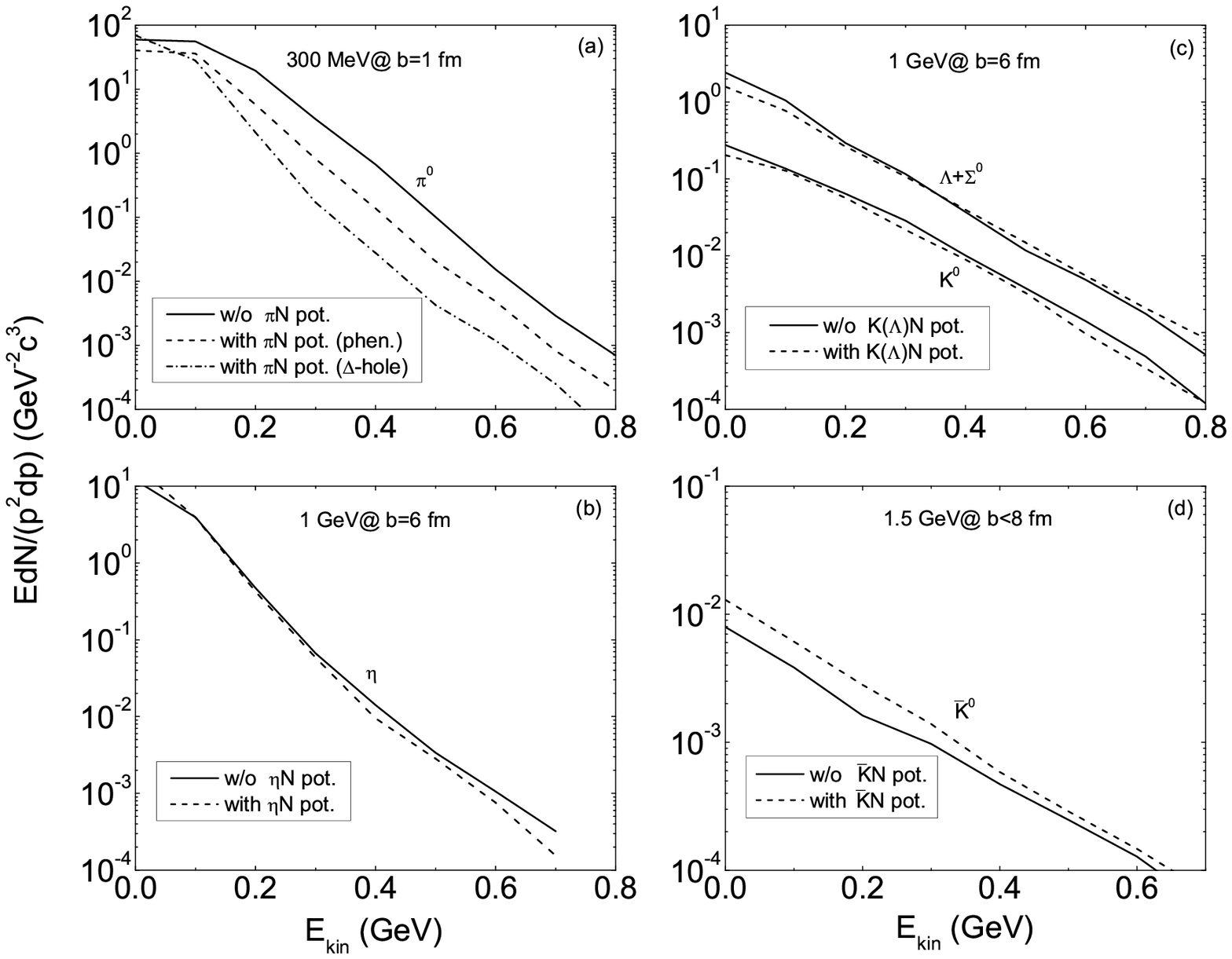}
\caption{\label{fig:wide} Inclusive spectra of $\pi^{0}$, $\eta$, $K^{0}$, $\overline{K}^{0}$ and neutral hyperons ($\Lambda$+$\Sigma^{0}$) in $^{197}$Au+$^{197}$Au collisions without (open symbols) and with (full symbols) the mean-field potentials, respectively.}
\end{figure*}

The in-medium effects of strange particles have been investigated from the K$^{-}$/K$^{+}$ spectrum in heavy-ion collisions and in proton induced reactions both in experiments \cite{La99,Fo03,Sc06} and in theories \cite{Li97,Fu06,Ha12,Fe13,Ca99}. A deeply attractive $K^{-}N$ potential being the value of -110$\pm$15 MeV was obtained at saturation density and weakly repulsive $K^{+}N$ potential has been concluded to be 25$\pm$10 MeV from heavy-ion collisions. Shown in Fig. 11 is a comparison of the K$^{-}$/K$^{+}$ ratio with and without the kaon(antikaon)-nucleon potentials in the $^{58}$Ni+$^{58}$Ni reaction as a function of transverse mass ($m_{t}=\sqrt{p_{t}^{2}+m_{0}^{2}}$ with $p_{t}$ being the transverse momentum and the mass of kaon (antikaon) $m_{0}$). The reduction of the threshold energies increases the production cross sections of antikaons. Furthermore, the $\overline{K}N$ potential enhances the low-momentum (kinetic energy) $K^{-}$ production. The $K^{-}/K^{+}$ ratio could be sensitive observable to extract the in-medium potentials. More information of the in-medium effects is also investigated from the total multiplicity of particles produced in central $^{40}$Ca+$^{40}$Ca collisions as shown in Fig. 12. It has been shown that the mean-field potentials of particles in nuclear medium have significant contributions on the total number of particle production. Specifically, the $\pi$N potential increases the collision probabilities of pions and nucleons, which leads to a reduction of pion numbers, in particular near threshold energies. Moreover, the $\eta$ and strange particles ($K^{0}$, $\overline{K}^{0}$, $\Lambda$, $\Sigma^{0}$) are enhanced owing to the $\pi$ induced reactions, such as $\pi N\rightarrow N\ast(1535)$, $\pi N\rightarrow KY$ and $\pi N\rightarrow NK\overline{K}$. The available data from FOPI collaboration for pions \cite{Re07} and from TAPS collaboration for etas \cite{Av03} can be well reproduced. It should be mentioned that the inclusion of the KN potential in the model leads to about 30$\%$ reduction of the total kaon yields in the subthreshold domain \cite{Fe13}. The competition of the $\pi (\eta)$N and KN potentials results in a bit of increase of kaon production because of the contribution of $\pi (\eta)N\rightarrow KY$.

\begin{figure*}
\includegraphics{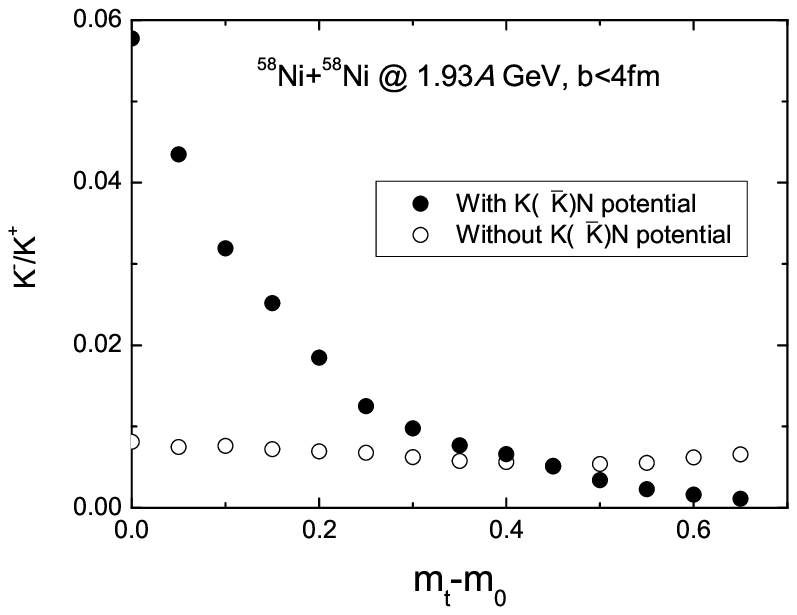}
\caption{\label{fig:wide} Ratio of K$^{-}$/K$^{+}$ as a function of transverse mass in collisions of $^{58}$Ni+$^{58}$Ni at the beam energy of 1.93\emph{A} GeV within the centrality of b$<$4 fm.}
\end{figure*}

\begin{figure*}
\includegraphics{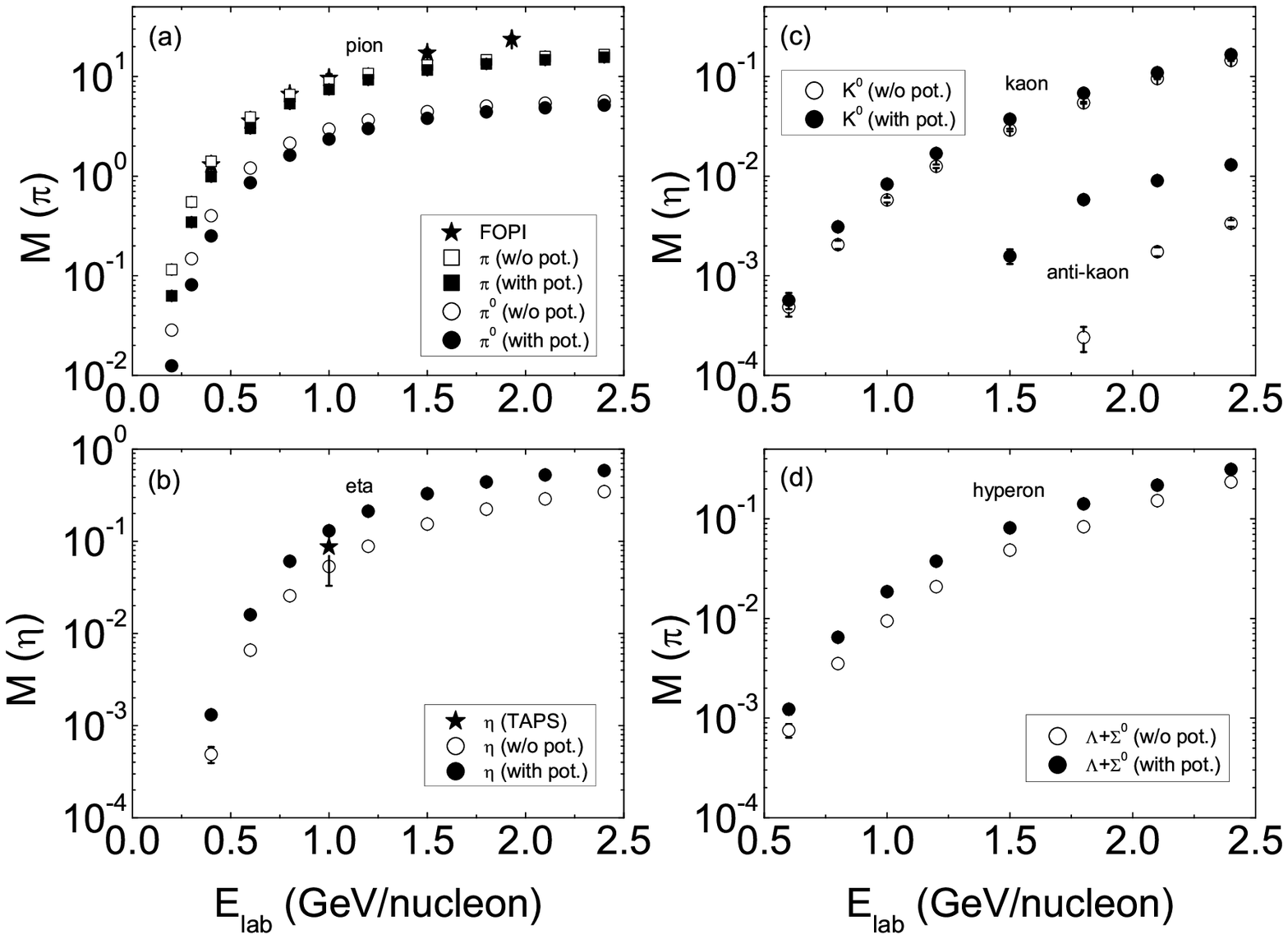}
\caption{\label{fig:wide} Excitation functions of neutral particles produced in central $^{40}$Ca+$^{40}$Ca collisions without (open symbols) and with (full symbols) the mean-field potentials, respectively. The available data from FOPI collaboration for pions \cite{Re07} and from TAPS collaboration for etas \cite{Av03} are shown by stars.}
\end{figure*}

The studies of the transverse flows of nucleons, light fragments, pions and strange particles in heavy-ion collisions have been motivated a lot of issues, such as the symmetry energy, in-medium NN cross section, optical potentials of particles in nuclear matter etc \cite{Fe13,Fe12b}. To investigate neutral particles production in the reaction plane and impacts of the optical potentials on particle dynamics, we computed the rapidity distributions of transverse flows for neutral particles ($\pi^{0}$, $\eta$, $K^{0}$ and $\Lambda$) produced in the peripheral $^{58}$Ni+$^{58}$Ni collisions (b=7 fm) at the incident energy of 1.93\emph{A} GeV as shown in Fig. 13. The attractive $\eta$N potential enlarges the transverse emission of $\eta$ in comparison to the in-vacuum case in heavy-ion collisions. Similar effect is also found for the $\Lambda$ production because of weakly attractive interaction between hyperons and nucleons below the baryon densities of 2.5$\rho_{0}$. Almost isotropic emission and a 'clock' rotation for $K^{0}$ production take place with the KN potential. The $\pi$N potential enhances the absorption of $\pi^{0}$ by surrounding nucleons and even appears the evidence of antiflow in comparison with the proton flow.

\begin{figure*}
\includegraphics{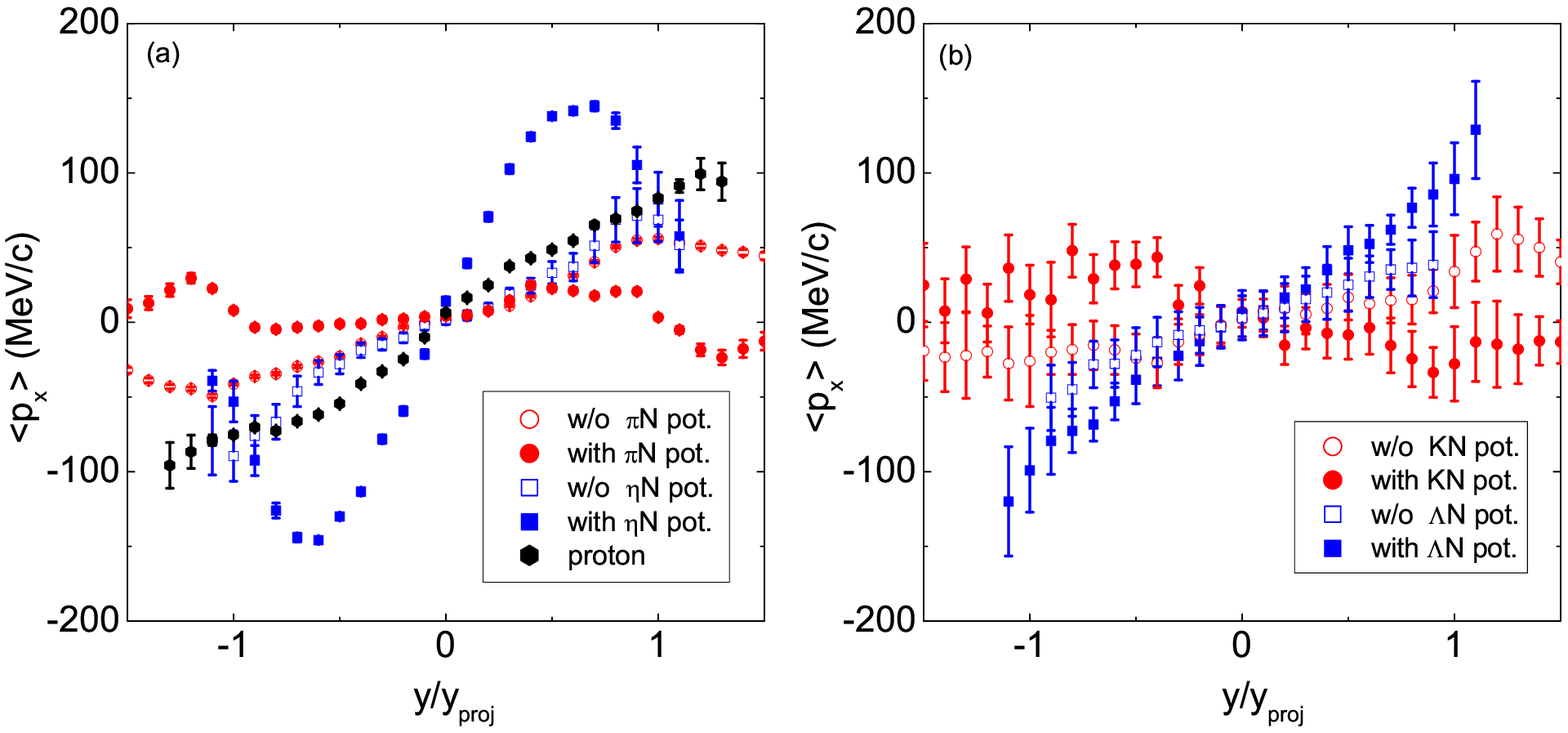}
\caption{\label{fig:wide} Rapidity distributions of transverse flows of neutral particles produced in the $^{58}$Ni+$^{58}$Ni reaction at an incident energy of 1.93\emph{A} GeV with an impact parameter b=7 fm.}
\end{figure*}

\section{IV. Conclusions}

The in-medium and isospin effects of pseudoscalar mesons and hyperons produced in heavy-ion collisions near threshold energies have been investigated within an isospin and momentum dependent hadron-transport model (LQMD). The in-medium potentials and corrections on threshold energies on particle production are of importance on particle transportation and distribution at freeze-out in phase space. The isospin related pion-nucleon potential reduces the total pion production, but enhances the $\pi^{-}/\pi^{+}$ ratio, in particular in the domain of subthreshold energies. The transverse flows, invariant spectra, rapidity and transverse momentum distributions in heavy-ion collisions are promising observables in extracting the in-medium properties of pions in dense nuclear matter. Experiments in the near future such as HIRFL-CSR (Lanzhou), RIKEN-SAMURAI in Japan etc, are expected for extracting the high-density symmetry energy, pion-nucleon potential, in-medium properties of $\Delta(1232)$.

The dynamics of etas, kaons, antikaons and hyperons in heavy-ion collisions is influenced by the in-medium potentials, in particular at the high-momentum tails. The transverse flow structure, invariant spectra, transverse mass spectra of $K^{-}/K^{+}$, rapidity and transverse momentum distributions are related to the mean-field potentials, in particular in the domain of subthreshold energies. The stiffness of symmetry energy weakly impacts the eta distribution in phase space.

\section{Acknowledgements}

We would like to thank Lie-Wen Chen, Maria Colonna, Massimo Di Toro, Alexei Larionov, Bao-An Li, Che Ming Ko, Ulrich Mosel, Hermann Wolter for fruitful discussions. This work was supported by the Major State Basic Research Development Program in China (No. 2014CB845405 and No. 2015CB856903), the National Natural Science Foundation of China Projects (Nos 11175218 and U1332207), the Youth Innovation Promotion Association of Chinese Academy of Sciences, and Yuncheng University Research Project (No. YQ-2014014).


\begin{thebibliography}{99}

\bibitem{Gi95} B. E. Gibson and E. V. Hungerford III, Phys. Rep. \textbf{257}, 349 (1995).
\bibitem{Fr07} E. Friedman and A. Gal, Phys. Rep. \textbf{452}, 89 (2007).
\bibitem{To11} M. Di Toro \emph{et al.}, Phys. Rev. C \textbf{83}, 014911 (2011).
\bibitem{Li95} G. Q. Li, C. M. Ko, and B. A. Li, Phys. Rev. Lett. \textbf{74}, 235 (1995); G. Q. Li and C. M. Ko, Nucl. Phys. A \textbf{594}, 460 (1995).
\bibitem{Li97} G. Q. Li, C. H. Lee, and G. E. Brown, Nucl. Phys. A \textbf{625}, 372 (1997); G. Q. Li and G. E. Brown \textbf{636}, 487 (1998).
\bibitem{Ca99} W. Cassing and E. L. Bratkovskaya, Phys. Rep. \textbf{308}, 65 (1999).
\bibitem{Fu06} C. Fuchs, Prog. Part. Nucl. Phys. \textbf{56}, 1 (2006).
\bibitem{Ha12} C. Hartnack, H. Oeschler, Y. Leifels, E. Bratkovskaya, and J. Aichelin, Phys. Rep. \textbf{510}, 119 (2012).
\bibitem{Li00} B. A. Li, Phys. Rev. Lett. \textbf{85}, 4221 (2000); Z. Q. Feng, Phys. Lett. B \textbf{707}, 83 (2012).
\bibitem{Ai85} J. Aichelin and C. M. Ko, Phys. Rev. Lett. \textbf{55}, 2661 (1985).
\bibitem{St01} C. Sturm \emph{et al.} (KaoS Collaboration), Phys. Rev. Lett. \textbf{86}, 39 (2001).
\bibitem{Li95b} G. Q. Li and C. M. Ko, Phys. Lett. B \textbf{349}, 405 (1995).
\bibitem{Fu01} C. Fuchs, A. Faessler, E. Zabrodin, and Y. M. Zheng, Phys. Rev. Lett. \textbf{86}, 1974 (2001).
\bibitem{Ha06} C. Hartnack, H Oeschler, and J. Aichelin, Phys. Rev. Lett. \textbf{96}, 012302 (2006).
\bibitem{Fe11} Z. Q. Feng, Phys. Rev. C \textbf{83}, 067604 (2011); Z. Q. Feng, Nucl. Phys. Rev. \textbf{31}, 326 (2014).
\bibitem{Li02} B. A. Li, Phys. Rev. Lett. \textbf{88}, 192701 (2002); B. A. Li, L. W. Chen, and C. M. Ko, Phys. Rep. \textbf{464}, 113 (2008).
\bibitem{Li05} Q. Li, Z. Li, E. Zhao, and R.K. Gupta, Phys. Rev. C \textbf{71}, 054907 (2005).
\bibitem{Fe06} G. Ferini, T. Gaitanos, M. Colonna, M. DiToro, and H. H. Wolter, Phys. Rev. Lett. \textbf{97}, 202301 (2006).
\bibitem{To10} M. Di Toro, V Baran, M Colonna, and V Greco, J. Phys. G: Nucl. Part. Phys. \textbf{37}, 083101 (2010).
\bibitem{Pr10} V. Prassa, T. Gaitanos, G. Ferini \emph{et al.}, Nucl. Phys. A \textbf{832}, 88 (2010).
\bibitem{Fe10} Z. Q. Feng and G.M. Jin, Phys. Rev. C \textbf{82}, 044615 (2010); Z. Q. Feng, Phys. Rev. C \textbf{87}, 064605 (2013).
\bibitem{Yo08} G. C. Yong, B. A. Li, and L. W. Chen, Phys. Lett. B \textbf{661}, 82 (2008); G. C. Yong and B. A. Li, Phys. Lett. B \textbf{723}, 388 (2013).
\bibitem{Gi10} V. Giordano, M. Colonna, M. Di Toro, V. Greco, J. Rizzo, Phys. Rev. C \textbf{81}, 044611 (2010).
\bibitem{Fe12} Z. Q. Feng, Phys. Rev. C \textbf{84}, 024610 (2011); Nucl. Phys. A \textbf{878}, 3 (2012); Nucl. Sci. Tech., \textbf{24}, 050504 (2013).
\bibitem{Xi09} Z. G. Xiao, B. A. Li, L. W. Chen, G. C. Yong, and M. Zhang, Phys. Rev. Lett. \textbf{102}, 062502 (2009).
\bibitem{Fe10b} Z. Q. Feng and G. M. Jin, Phys. Lett. B \textbf{683}, 140 (2010).
\bibitem{Xi13} W. J. Xie, J. Su, L. Zhu, and F. S. Zhang, Phys. Lett. B \textbf{718}, 1510 (2013).
\bibitem{Re07} W. Reisdorf \emph{et al.} (FOPI Collaboration), Nucl. Phys. A \textbf{781}, 459 (2007).
\bibitem{Fe05} G. Ferini, M. Colonna, T. Gaitanos, and M. Di Toro, Nucl. Phys. A \textbf{762}, 147 (2005).
\bibitem{So15} T. Song and C. M. Ko, Phys. Rev. C \textbf{91}, 014901 (2015).
\bibitem{Xi93} L. Xiong, C. M. Ko, and V. Koch, Phys. Rev. C \textbf{47}, 788 (1993).
\bibitem{Fu97} C. Fuchs, L. Sehn, E. Lehmann, J. Zipprich, and A. Faessler, Phys. Rev. C \textbf{55}, 411 (1997).
\bibitem{Ho14} J. Hong and P. Danielewicz, Phys. Rev. C \textbf{90}, 024605 (2014).
\bibitem{Fe09} Z. Q. Feng and G. M. Jin, Chin. Phys. Lett. \textbf{26}, 062501 (2009); Phys. Rev. C \textbf{82}, 057901 (2010).
\bibitem{Fe14} Z. Q. Feng, H. Lenske, Phys. Rev. C \textbf{89}, 044617 (2014); Z. Q. Feng, Nucl. Sci. Tech., \textbf{26}, S20512 (2015).
\bibitem{He15} O. Hen, B. A. Li, W. J. Guo, L. B. Weinstein, and E. Piasetzky, Phys. Rev. C \textbf{91}, 025803 (2015).
\bibitem{La09} A. B. Larionov, I. A. Pshenichnov, I. N. Mishustin, and W. Greiner, Phys. Rev. C \textbf{80}, 021601 (2009).
\bibitem{Co82} J. C\^{o}nt\'{e}, M. Lacombe, B. Loiseau, B. Moussallam, and R. Vinh Mau, Phys. Rev. Lett. \textbf{48}, 1319 (1982).
\bibitem{Ga87} C. Gale and J. Kapusta, Phys. Rev. C \textbf{35}, 2107 (1987).
\bibitem{Br75} G. E. Brown and W. Weise, Phys. Rep. \textbf{22}, 279 (1975); B. Friemann, V. P. Pandharipande, and Q. N. Usmani, Nucl. Phys. A \textbf{372}, 483 (1981).
\bibitem{Xu10} J. Xu, C. M. Ko, and Y. Oh, Phys. Rev. C \textbf{81}, 024910 (2010); J. Xu, L. W. Chen, C. M. Ko, B. A. Li, and Y. G. Ma, Phys. Rev. C \textbf{87}, 067601 (2013).
\bibitem{Ni08} P. Z. Ning, \emph{Strangeness Nuclear Physics (in chinese)} (Science Press, 2008); X. H. Zhong, G. X. Peng, Lei Li, and P. Z. Ning, Phys. Rev. C \textbf{73}, 015205 (2006).
\bibitem{Ch91} H. C. Chiang, E. Oset, and L. C. Liu, Phys. Rev. C \textbf{44}, 738 (1991); T. Waas and W. Weise, Nucl. Phys. A \textbf{625}, 287 (1997); K. Tsushima, D. H. Lu, A. W. Thomas, and K. Saito, Phys. Lett. B \textbf{443}, 26 (1998); T. Inoue and E. Oset, Nucl. Phys. A \textbf{710}, 354 (2002).
\bibitem{Li86} L. C. Liu and Q. Haider, Phys. Rev. C \textbf{34}, 1845 (1986).
\bibitem{Ka86} D. B. Kaplan and A. E. Nelson, Phys. Lett. B \textbf{175}, 57 (1986).
\bibitem{Sc97} J. Schaffner-Bielich, I.N. Mishustin, and J. Bondorf, Nucl. Phys. A \textbf{625}, 325 (1997).
\bibitem{Fe13} Z. Q. Feng, Nucl. Phys. A \textbf{919}, 32 (2013); Z. Q. Feng, W. J. Xie, and G. M. Jin, Phys. Rev. C \textbf{90}, 064604 (2014).
\bibitem{Zh04} Y. M. Zheng, C. Fuchs, A. Faessler \emph{et al.}, Phys. Rev. C \textbf{69}, 034907 (2004).
\bibitem{Sh92} E. Shuryak and V. Thorsson, Nucl. Phys. A \textbf{536}, 739 (1992); A. Sibirtsev and W. Cassing, Nucl. Phys. A \textbf{641}, 476 (1998).
\bibitem{Hu94} S. Huber and J. Aichelin, Nucl. Phys. A \textbf{573}, 587 (1994).
\bibitem{Fe12b} Z. Q. Feng, Phys. Rev. C \textbf{85}, 014604 (2012).
\bibitem{Ts99} K. Tsushima, A. Sibirtsev, A. W. Thomas, and G. Q. Li, Phys. Rev. C \textbf{59}, 369 (1999).
\bibitem{Ts94} K. Tsushima, S.W. Huang, and A. Faessler, Phys. Lett. B \textbf{337}, 245 (1994); J. Phys. G \textbf{21}, 33 (1995).
\bibitem{Cu84} J. Cugnon and R.M. Lombard, Nucl. Phys. A \textbf{422}, 635 (1984).
\bibitem{Ca97} W. Cassing \emph{et al.}, Nucl. Phys. A \textbf{614}, 415 (1997).
\bibitem{Cu90} J. Cugnon, P. Deneye, and J. Vandermeulen, Phys. Rev. C \textbf{41}, 1701 (1990).
\bibitem{Li01} B. A. Li, A. T. Sustich, B. Zhang, and C. M. Ko, Int. J. Mod. Phys. E \textbf{10}, 1 (2001).
\bibitem{La99} F. Laue \emph{et al.}, Phys. Rev. Lett. \textbf{82}, 1640 (1999).
\bibitem{Fo03} A. F\"{o}rster \emph{et al.} (KaoS Collaboration), Phys. Rev. Lett. \textbf{91}, 152301 (2003).
\bibitem{Sc06} W. Scheinast \emph{et al.}, Phys. Rev. Lett. \textbf{96}, 072301 (2006).
\bibitem{Av03} R. Averbeck, R. Holzmann, V. Metag, and R. S. Simon, Phys. Rev. C \textbf{67}, 024903 (2003).

\end{thebibliography}
\end{document}